\documentstyle[epsf,11pt]{article}
\newcommand{\pr}{{Phys.\ Rev.\/}}

\title{On the influence of colour magnetic currents on the confining
properties of $SU(3)$ lattice gauge theory\thanks{Supported in part by
Fonds zur F\"orderung der wissenschaftlichen Forschung, P11387-PHY}} 

\author{Peter Skala, Manfried Faber and Martin Zach\\
Institut f\"ur Kernphysik, Technische Universit\"at Wien\\
A--1040 Vienna, Austria}

\date{}

\begin{document}

\maketitle

\begin{abstract}
We modify the standard Wilson action of $SU(3)$ lattice gauge theory
by adding an extra term which suppresses colour magnetic currents. We
present numerical results of simulations at zero and finite
temperature and show that colour magnetic currents strongly influence
the confining properties of $SU(3)$ lattice gauge theory.

\end{abstract}

\section{Introduction}

The mechanism which leads to permanent confinement of quarks at low
temperatures is an interesting and still open question in gauge
theories. A promising conjecture which could explain the existence of
a non-zero string tension is the hypothesis that the QCD vacuum
behaves dually to a superconductor \cite{mandel}. In this model colour
magnetically charged particles condense in the ground state of QCD and
squeeze the colour electric field between a pair of static quarks into
a narrow flux tube leading to a confining potential. A possible
realization of the dual superconductor picture in non-Abelian $SU(N)$
gauge theories was proposed by 't~Hooft in the framework of Abelian
projection \cite{thooft}. In this approach, the physically relevant
degrees of freedom are identified by fixing the gauge up to the
largest Abelian subgroup $U(1)^{N-1}$ of the $SU(N)$ gauge
group. After gauge fixing one is left with an effective Abelian theory
which contains in addition to photons magnetic monopoles. The
monopoles arise as singularities in the gauge fixing condition and
their condensation is expected to be crucial for the existence of a
non-zero string tension. There is a lot of support for the Abelian
projection approach from numerical results of lattice simulations
\cite{poli}. For instance, the Abelian gauge fields obtained after
gauge fixing carry almost the whole asymptotic string tension
\cite{suzu}. This observation is referred to as Abelian
dominance. However, most of the numerical results are 
dependent on the choice of gauge which fixes the non-Abelian degrees
of freedom \cite{poli}, and a dual superconductor is convincingly 
observed only in one particular gauge, the so-called maximal Abelian
gauge \cite{kron}. The question arises whether this apparent ambiguity
in the gauge fixing procedure can be avoided by investigating the dual
superconductor picture in terms of the original non-Abelian degrees of
freedom.

A first attempt in this direction was made in ref.~\cite{skala} where
dual superconductivity in $SU(3)$ lattice gauge theory was studied
numerically in a gauge invariant formulation. An operator for the
magnetic current was introduced allowing the determination of its curl
in the presence of a static quark antiquark pair without need of
fixing the gauge. At finite temperature the result of a numerical
simulation is the following \cite{skala}: In the confined phase the
curl of the magnetic current is proportional to the electric field
indicating the validity of a dual London relation. This seems to be in
agreement with the dual superconductor picture of confinement. In the
deconfined phase the curl of the magnetic current vanishes and the
electric field shows Coulomb-like behaviour. At least qualitatively
the results presented in \cite{skala} agree with the results reported
in ref.~\cite{haymaker} which were obtained in the maximal Abelian gauge.

In this letter, we continue the gauge invariant investigation of the
dual superconductor picture of confinement and study the role of the
magnetic current operator introduced in ref.~\cite{skala} in the pure
gluonic vacuum. We introduce
a chemical potential $\lambda$ and add an extra term to the standard
gauge field action suppressing large magnetic currents. By calculating
the string tension and the critical coupling for the finite
temperature phase transition we will show that a suppression of
magnetic currents strongly influences the confining properties of the
considered $SU(3)$ theory.

\section{The Model}

We consider pure $SU(3)$ lattice gauge theory on a four-dimensional
Euclidean lattice of spacing $a$ with periodic boundary conditions in
space and time direction. An action convenient for many purposes is
the standard Wilson action \cite{wilson} which for the $SU(3)$ gauge
group reads

\begin{equation}
\label{waction}
S_W \; = \; \beta \sum_{x,\mu < \nu} \left( 1 - \frac{1}{3} \,
\mbox{Re} \, \mbox{Tr} \, U_{\mu\nu}(x) \right), 
\end{equation}
where $\beta = 6/g^2$ is the inverse coupling and $U_{\mu\nu}(x)$ is
the product of link variables $U_\mu \in SU(3)$ around an elementary
plaquette in $\mu\nu$-direction at lattice site $x$: 
\begin{equation}
\label{plaquette}
U_{\mu\nu}(x) \; = \;
U_{\mu}(x)U_{\nu}(x+\hat{\mu})U_{\mu}^\dagger(x+\hat{\nu})U_\nu^\dagger(x). 
\end{equation}
According to a derivation of Gauss' law on the lattice \cite{skala}
the field strength $F^x_{\mu\nu}(x)$ corresponding to the action
(\ref{waction}) is given by the expression
\begin{equation}
\label{fstrength}
ga^2 \, F_{\mu\nu}^x(x) \; = \; {\cal F}_{\mu\nu}^x(x) = \;
\frac{1}{2i}\left( U_{\mu\nu}(x) - U_{\mu\nu}^\dagger(x)
\right)_{(tl)}, \;\;\;\;\;\;\;\;\; (tl) \equiv \mbox{traceless}. 
\end{equation}
The upper index $x$ in the notation of $F^x_{\mu\nu}(x)$ states the
local colour coordinate system in which the field strength is
measured. It is changed to a system defined at an arbitrary lattice
site $x'$ by performing a parallel transport of the field strength
along a Schwinger line connecting the site $x'$ with the original site
$x$.

Let us now introduce the quantity $J^x_{m,\mu}$ \cite{skala} given by
the expression
\begin{equation}
\label{mcur}
J^x_{m,\mu}(x) \, \equiv \, -\frac{1}{2} \epsilon_{\mu\nu\rho\sigma}
D_\nu {\cal F}_{\rho\sigma}^x(x),
\end{equation}
where we use the discretized covariant derivative of the field
strength 
\begin{equation}
\label{cderive}
D_{\nu}{\cal F}_{\rho\sigma}^x(x) \, \equiv \,{\cal
F}_{\rho\sigma}^x(x+\hat{\nu}) - {\cal F}_{\rho\sigma}^x(x) \, = \,
U_\nu(x) {\cal
F}_{\rho\sigma}^{x+\hat{\nu}}(x+\hat{\nu})U_\nu^\dagger(x) - {\cal
F}_{\rho\sigma}^x(x). 
\end{equation}
On the lattice definition (\ref{mcur}) has a simple geometrical
interpretation. Each component of $J_{m,\mu}$ corresponds to a
three-dimensional cube built of six plaquettes measuring the flux out
of the cube. We therefore call $J_{m,\mu}$ colour magnetic
current. The covariant derivative in (\ref{mcur}) guarantees all
contributions to the total flux out of the cube to be taken in the
colour coordinate system at $x$. As can be easily shown \cite{skala}
the current (\ref{mcur}) fulfills a conservation law in terms of
covariant derivatives.
 
In ref.~\cite{skala} we already discussed the definition of the
quantity $J_{m,\mu}$ concerning the validity of the Bianchi
identity. We think it is necessary to extend this discussion and add
at this point some further comments concerning especially the general
form of the Bianchi identity in a lattice formulation of $SU(3)$ gauge
theory. In a continuum formulation the right hand side of (\ref{mcur})
corresponds to the Bianchi identity and the current $J_{m,\mu}$
vanishes. On a lattice with finite lattice spacing $a$, however,
definition (\ref{mcur}) may not be identified with a lattice version
of the Bianchi identity. This can be understood as follows: As was
shown in ref.~\cite{batrouni} the partition function $Z$ of $SU(3)$
lattice gauge theory expressed in terms of link variables $U_{\mu}(x)
\in SU(3)$
\begin{equation}
\label{zpart}
Z = \int { \cal D }[U_{\mu}(x)] \exp (-S_W[U_{\mu}])
\end{equation}
can be rewritten as an integral over the plaquette variables
$U_{\mu\nu}(x) \in SU(3)$. The lattice Bianchi identity then
automatically arises as the argument of a $\delta$-function in the
integrand of the partition function. It essentially constrains the
product of plaquettes $U_{\mu\nu}$ covering a three-dimensional cube
to the unit matrix in order to reduce the degrees of freedom (6
plaquettes per lattice site) to the original number (4 links per
lattice site). On a four-dimensional lattice there are four types of
three-dimensional cubes and thus four Bianchi identities. We want to
emphasize that if a theory is expressed in terms of link variables the
lattice Bianchi identities for the plaquette variables $U_{\mu\nu}$
are automatically fulfilled. In the case of an Abelian theory like
$U(1)$ gauge theory the lattice Bianchi identity can be formulated
also as a constraint for the plaquette phases. The sum of the phases
$\theta_{\mu\nu} \in (-\pi,\pi]$ of the six plaquette variables
$U_{\mu\nu}=e^{i\theta_{\mu\nu}} \in U(1)$ measuring the flux out of a
three-dimensional cube must be an integer multiple of $2 \pi$. If the
integer is non-zero, one speaks of a topological excitation called
magnetic monopole \cite{degrand}. However, if one identifies the flux
through a plaquette according to (\ref{fstrength}) with $\sin
\theta_{\mu\nu}$ \cite{zach}, the total flux out of a
three-dimensional cube is in general non-zero. The last statement is
also true in the case of a non-Abelian theory without being in
contradiction to the lattice Bianchi identity. We conclude that the
sum of algebra elements ${\cal F}_{\mu\nu}$ on the right hand side of
(\ref{mcur}) is in general non-zero although the product of the
corresponding group elements $U_{\mu\nu}$ is - according to the
Bianchi identity - equal to the unit matrix. 

It is the purpose of this paper to study the influence of the magnetic
current $J_{m,\mu}$ on the confining properties of the considered
$SU(3)$ lattice gauge theory. To be more precise, we define the
following gauge invariant quantity 
\begin{equation}
\label{jnorm}
| J_{m,\mu}(x) | = \sqrt{2 \; \mbox{Tr} \; \left( J^x_{m,\mu}(x)
  \right)^2} 
\end{equation}
and call it the length in colour space of the magnetic current defined
in (\ref{mcur}). Large values of $|J_{m,\mu}(x) |$ correspond to large
fluctuations of the plaquettes covering the surface of a
three-dimensional cube. We introduce a parameter $\lambda$ and modify
the Wilson action (\ref{waction}) in the following way 
\begin{equation}
\label{maction}
S = S_W + \lambda \sum_{x,\mu} |J_{m,\mu}(x) |.
\end{equation}
The sum extends over all three-dimensional cubes on the lattice. The
parameter $\lambda$ plays the role of a chemical potential for the
magnetic current $J_{m,\mu}$. If $\lambda$ is chosen to be larger than
zero, the magnetic current $J_{m,\mu}$ will be suppressed. In
ref.~\cite{schrader} the action (\ref{maction}) with $\lambda =
\infty$ and with $J_{m,\mu}$ being the $U(1)$ monopole currents
\cite{degrand} was used to study four-dimensional $U(1)$ lattice gauge
theory with monopoles removed.  

\section{Results}

In order to study the influence of magnetic currents $J_{m,\mu}$ on
physical quantities we performed simulations using the action
(\ref{maction}) with zero and non-zero chemical potential
$\lambda$. Calculations were done on symmetric lattices with size
$8^4$ and $12^4$ and on asymmetric lattices with temporal extent
$N_{\tau}=4$ and spatial extent of $N_{\sigma}=8$ and $12$. The case
$\lambda=0$, i.e.~the standard Wilson action, was simulated by using
an overrelaxed pseudo heatbath algorithm. To generate gauge field
configurations at non-zero chemical potential $\lambda$ we implemented
a Metropolis algorithm. Because of the additional term in the action
twelve three-dimensional cubes corresponding to colour magnetic
currents $J_{m,\mu}$ have to be taken into account besides the six
plaquettes in the update of a single link. This procedure turned out
to be very time consuming. We therefore decided to measure the
physical quantities after every Metropolis sweep; this leads to large
autocorrelation times especially in the critical regions of the finite
temperature phase transition. As a consequence a large number of
measurements up to $10^5$ had to be taken into account in order to get
a reliable estimate of the considered physical quantities. To estimate
the errorbars of primary quantities such as Polyakov and Wilson loops
the standard procedure of calculating the integrated autocorrelation
time was used. For the error analysis of secondary quantities such as
Creutz ratios a simple extension of the fundamental jackknife formula
\cite{gottlieb} was applied.
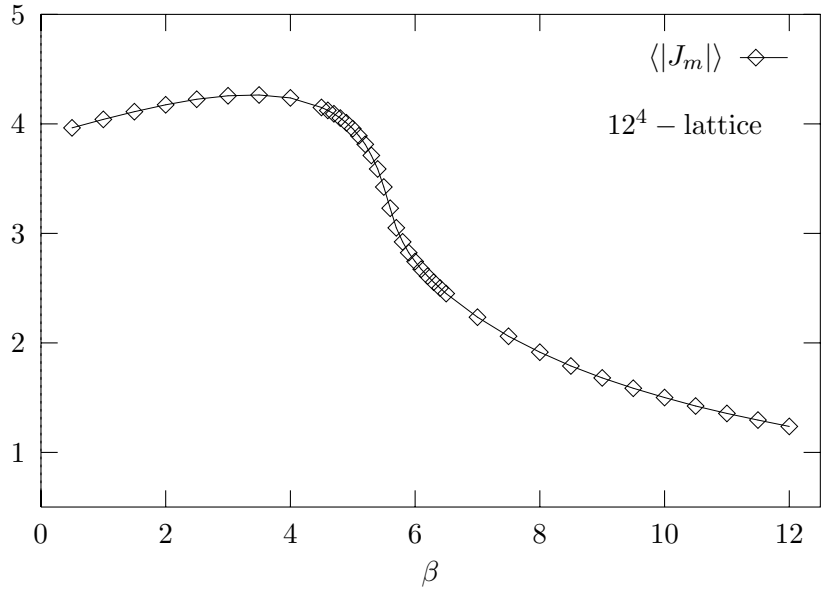
\begin{figure}
\centerline{
\setlength{\unitlength}{0.1bp}
\special{!
/gnudict 40 dict def
gnudict begin
/Color false def
/Solid false def
/gnulinewidth 2.000 def
/vshift -33 def
/dl {10 mul} def
/hpt 31.5 def
/vpt 31.5 def
/M {moveto} bind def
/L {lineto} bind def
/R {rmoveto} bind def
/V {rlineto} bind def
/vpt2 vpt 2 mul def
/hpt2 hpt 2 mul def
/Lshow { currentpoint stroke M
  0 vshift R show } def
/Rshow { currentpoint stroke M
  dup stringwidth pop neg vshift R show } def
/Cshow { currentpoint stroke M
  dup stringwidth pop -2 div vshift R show } def
/DL { Color {setrgbcolor Solid {pop []} if 0 setdash }
 {pop pop pop Solid {pop []} if 0 setdash} ifelse } def
/BL { stroke gnulinewidth 2 mul setlinewidth } def
/AL { stroke gnulinewidth 2 div setlinewidth } def
/PL { stroke gnulinewidth setlinewidth } def
/LTb { BL [] 0 0 0 DL } def
/LTa { AL [1 dl 2 dl] 0 setdash 0 0 0 setrgbcolor } def
/LT0 { PL [] 0 1 0 DL } def
/LT1 { PL [4 dl 2 dl] 0 0 1 DL } def
/LT2 { PL [2 dl 3 dl] 1 0 0 DL } def
/LT3 { PL [1 dl 1.5 dl] 1 0 1 DL } def
/LT4 { PL [5 dl 2 dl 1 dl 2 dl] 0 1 1 DL } def
/LT5 { PL [4 dl 3 dl 1 dl 3 dl] 1 1 0 DL } def
/LT6 { PL [2 dl 2 dl 2 dl 4 dl] 0 0 0 DL } def
/LT7 { PL [2 dl 2 dl 2 dl 2 dl 2 dl 4 dl] 1 0.3 0 DL } def
/LT8 { PL [2 dl 2 dl 2 dl 2 dl 2 dl 2 dl 2 dl 4 dl] 0.5 0.5 0.5 DL } def
/P { stroke [] 0 setdash
  currentlinewidth 2 div sub M
  0 currentlinewidth V stroke } def
/D { stroke [] 0 setdash 2 copy vpt add M
  hpt neg vpt neg V hpt vpt neg V
  hpt vpt V hpt neg vpt V closepath stroke
  P } def
/A { stroke [] 0 setdash vpt sub M 0 vpt2 V
  currentpoint stroke M
  hpt neg vpt neg R hpt2 0 V stroke
  } def
/B { stroke [] 0 setdash 2 copy exch hpt sub exch vpt add M
  0 vpt2 neg V hpt2 0 V 0 vpt2 V
  hpt2 neg 0 V closepath stroke
  P } def
/C { stroke [] 0 setdash exch hpt sub exch vpt add M
  hpt2 vpt2 neg V currentpoint stroke M
  hpt2 neg 0 R hpt2 vpt2 V stroke } def
/T { stroke [] 0 setdash 2 copy vpt 1.12 mul add M
  hpt neg vpt -1.62 mul V
  hpt 2 mul 0 V
  hpt neg vpt 1.62 mul V closepath stroke
  P  } def
/S { 2 copy A C} def
end
}
\begin{picture}(3600,2160)(0,0)
\special{"
gnudict begin
gsave
50 50 translate
0.100 0.100 scale
0 setgray
/Helvetica findfont 100 scalefont setfont
newpath
-500.000000 -500.000000 translate
LTa
480 251 M
0 1858 V
LTb
480 457 M
63 0 V
2874 0 R
-63 0 V
480 870 M
63 0 V
2874 0 R
-63 0 V
480 1283 M
63 0 V
2874 0 R
-63 0 V
480 1696 M
63 0 V
2874 0 R
-63 0 V
480 2109 M
63 0 V
2874 0 R
-63 0 V
480 251 M
0 63 V
0 1795 R
0 -63 V
950 251 M
0 63 V
0 1795 R
0 -63 V
1420 251 M
0 63 V
0 1795 R
0 -63 V
1890 251 M
0 63 V
0 1795 R
0 -63 V
2360 251 M
0 63 V
0 1795 R
0 -63 V
2830 251 M
0 63 V
0 1795 R
0 -63 V
3300 251 M
0 63 V
0 1795 R
0 -63 V
480 251 M
2937 0 V
0 1858 V
-2937 0 V
480 251 L
LT0
3114 1946 M
180 0 V
597 1681 M
118 32 V
117 29 V
118 26 V
117 21 V
118 13 V
117 3 V
118 -11 V
117 -36 V
24 -11 V
23 -13 V
24 -16 V
23 -18 V
24 -22 V
23 -27 V
24 -32 V
23 -42 V
24 -51 V
23 -68 V
24 -80 V
23 -74 V
24 -53 V
23 -41 V
24 -33 V
23 -29 V
24 -26 V
23 -23 V
24 -22 V
23 -21 V
118 -89 V
117 -72 V
118 -60 V
117 -52 V
118 -45 V
117 -39 V
118 -35 V
117 -32 V
118 -28 V
117 -25 V
118 -24 V
3174 1946 D
597 1681 D
715 1713 D
832 1742 D
950 1768 D
1067 1789 D
1185 1802 D
1302 1805 D
1420 1794 D
1537 1758 D
1561 1747 D
1584 1734 D
1608 1718 D
1631 1700 D
1655 1678 D
1678 1651 D
1702 1619 D
1725 1577 D
1749 1526 D
1772 1458 D
1796 1378 D
1819 1304 D
1843 1251 D
1866 1210 D
1890 1177 D
1913 1148 D
1937 1122 D
1960 1099 D
1984 1077 D
2007 1056 D
2125 967 D
2242 895 D
2360 835 D
2477 783 D
2595 738 D
2712 699 D
2830 664 D
2947 632 D
3065 604 D
3182 579 D
3300 555 D
stroke
grestore
end
showpage
}
\put(3054,1946){\makebox(0,0)[r]{$\langle | J_m | \rangle$}}
\put(3200,1700){\makebox(0,0)[r]{$12^4 - \mbox{lattice}$}}
\put(1948,0){\makebox(0,0){$\beta$}}
\put(3300,151){\makebox(0,0){12}}
\put(2830,151){\makebox(0,0){10}}
\put(2360,151){\makebox(0,0){8}}
\put(1890,151){\makebox(0,0){6}}
\put(1420,151){\makebox(0,0){4}}
\put(950,151){\makebox(0,0){2}}
\put(480,151){\makebox(0,0){0}}
\put(420,2109){\makebox(0,0)[r]{5}}
\put(420,1696){\makebox(0,0)[r]{4}}
\put(420,1283){\makebox(0,0)[r]{3}}
\put(420,870){\makebox(0,0)[r]{2}}
\put(420,457){\makebox(0,0)[r]{1}}
\end{picture}}
\caption{\label{jm}Expectation value of the magnetic current $|J_m|$
(\ref{jma}) as a function of the inverse coupling $\beta$. The
measurements were taken on a $12^4$-lattice. Errorbars are omitted
since they are smaller than the symbols.}
\end{figure}

We start with the presentation of results obtained on symmetric
lattices of size $8^4$ and $12^4$. Fig.~\ref{jm} shows the expectation
value of the magnetic current (\ref{jnorm}) averaged over the
space-time components and the four-dimensional lattice
\begin{equation}
\label{jma}
| J_m | = \frac{1}{4N_{\sigma}^4} \sum_{x,\mu}  |J_{m,\mu}(x)| 
\end{equation}
for a large range of $\beta$-values and zero chemical potential
$\lambda$. One can see a typical crossover of $\langle | J_m |
\rangle$ between the strong and weak coupling region. For $\beta$
going to infinity the magnetic current tends to zero displaying
decreasing fluctuations of the six plaquettes covering a
three-dimensional cube. In the strong coupling region the current
(\ref{jma}) surprisingly reaches its maximum at a finite value of
$\beta$ and slightly decreases for $\beta$ going to zero. It is an
interesting question whether the current density $\langle | J_m |
\rangle$ shows scaling behaviour which means $\langle | J_m |
\rangle \propto a^3(\beta)$, where $a(\beta)$ is the lattice spacing
given as a function of the bare coupling constant $\beta$ which can be
determined by the renormalization group equation. From the numerical
data presented in fig.~\ref{jm} it is clearly seen that there is no
scaling behaviour of $\langle | J_m | \rangle$. This may be explained
by observing that there are only positive contributions to the current
density $\langle | J_m | \rangle$. Quantum fluctuations which are not
of topological origin do not cancel. It should be emphazised that for
the correlation function measuring the curl of the magnetic current in
the vicinity of a static charge pair (see ref.~\cite{skala}) this
problem does not seem to occur. In this case, the magnetic current is
projected to the Polyakov line which fixes a direction in colour
space. Fluctuations corresponding to lattice artifacts contribute
equally with positive and negative sign and thus cancel, whereas
contributions of topological origin survive the averaging
process. Unfortunately, the situation is different in the case of the
operator $ | J_m | $ which measures the magnetic current density in
the gluonic vacuum. We will come back to the question of scaling of
this operator and its consequences to the continuum limit in the
conclusions of this letter.  

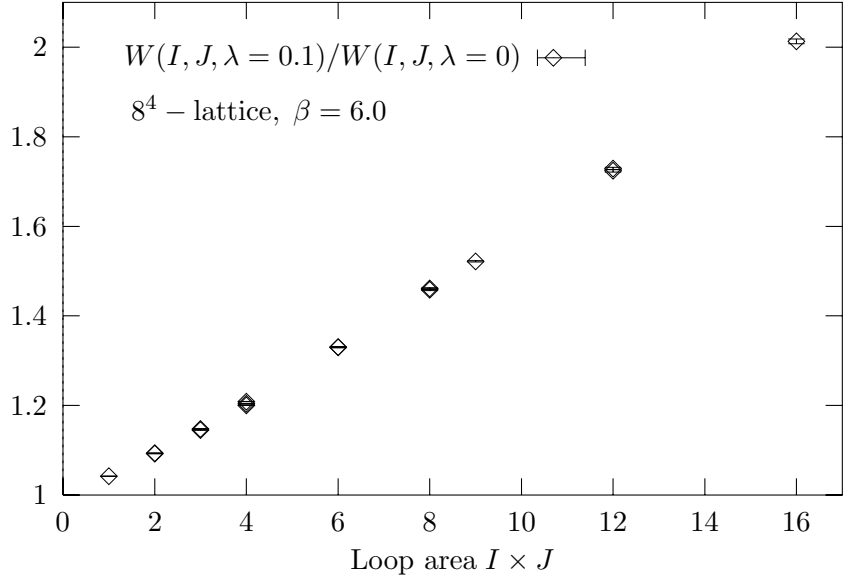
\begin{figure}
\centerline{
\setlength{\unitlength}{0.1bp}
\special{!
/gnudict 40 dict def
gnudict begin
/Color false def
/Solid false def
/gnulinewidth 2.000 def
/vshift -33 def
/dl {10 mul} def
/hpt 31.5 def
/vpt 31.5 def
/M {moveto} bind def
/L {lineto} bind def
/R {rmoveto} bind def
/V {rlineto} bind def
/vpt2 vpt 2 mul def
/hpt2 hpt 2 mul def
/Lshow { currentpoint stroke M
  0 vshift R show } def
/Rshow { currentpoint stroke M
  dup stringwidth pop neg vshift R show } def
/Cshow { currentpoint stroke M
  dup stringwidth pop -2 div vshift R show } def
/DL { Color {setrgbcolor Solid {pop []} if 0 setdash }
 {pop pop pop Solid {pop []} if 0 setdash} ifelse } def
/BL { stroke gnulinewidth 2 mul setlinewidth } def
/AL { stroke gnulinewidth 2 div setlinewidth } def
/PL { stroke gnulinewidth setlinewidth } def
/LTb { BL [] 0 0 0 DL } def
/LTa { AL [1 dl 2 dl] 0 setdash 0 0 0 setrgbcolor } def
/LT0 { PL [] 0 1 0 DL } def
/LT1 { PL [4 dl 2 dl] 0 0 1 DL } def
/LT2 { PL [2 dl 3 dl] 1 0 0 DL } def
/LT3 { PL [1 dl 1.5 dl] 1 0 1 DL } def
/LT4 { PL [5 dl 2 dl 1 dl 2 dl] 0 1 1 DL } def
/LT5 { PL [4 dl 3 dl 1 dl 3 dl] 1 1 0 DL } def
/LT6 { PL [2 dl 2 dl 2 dl 4 dl] 0 0 0 DL } def
/LT7 { PL [2 dl 2 dl 2 dl 2 dl 2 dl 4 dl] 1 0.3 0 DL } def
/LT8 { PL [2 dl 2 dl 2 dl 2 dl 2 dl 2 dl 2 dl 4 dl] 0.5 0.5 0.5 DL } def
/P { stroke [] 0 setdash
  currentlinewidth 2 div sub M
  0 currentlinewidth V stroke } def
/D { stroke [] 0 setdash 2 copy vpt add M
  hpt neg vpt neg V hpt vpt neg V
  hpt vpt V hpt neg vpt V closepath stroke
  P } def
/A { stroke [] 0 setdash vpt sub M 0 vpt2 V
  currentpoint stroke M
  hpt neg vpt neg R hpt2 0 V stroke
  } def
/B { stroke [] 0 setdash 2 copy exch hpt sub exch vpt add M
  0 vpt2 neg V hpt2 0 V 0 vpt2 V
  hpt2 neg 0 V closepath stroke
  P } def
/C { stroke [] 0 setdash exch hpt sub exch vpt add M
  hpt2 vpt2 neg V currentpoint stroke M
  hpt2 neg 0 R hpt2 vpt2 V stroke } def
/T { stroke [] 0 setdash 2 copy vpt 1.12 mul add M
  hpt neg vpt -1.62 mul V
  hpt 2 mul 0 V
  hpt neg vpt 1.62 mul V closepath stroke
  P  } def
/S { 2 copy A C} def
end
}
\begin{picture}(3600,2160)(0,0)
\special{"
gnudict begin
gsave
50 50 translate
0.100 0.100 scale
0 setgray
/Helvetica findfont 100 scalefont setfont
newpath
-500.000000 -500.000000 translate
LTa
480 251 M
0 1858 V
LTb
480 251 M
63 0 V
2874 0 R
-63 0 V
480 589 M
63 0 V
2874 0 R
-63 0 V
480 927 M
63 0 V
2874 0 R
-63 0 V
480 1264 M
63 0 V
2874 0 R
-63 0 V
480 1602 M
63 0 V
2874 0 R
-63 0 V
480 1940 M
63 0 V
2874 0 R
-63 0 V
480 251 M
0 63 V
0 1795 R
0 -63 V
826 251 M
0 63 V
0 1795 R
0 -63 V
1171 251 M
0 63 V
0 1795 R
0 -63 V
1517 251 M
0 63 V
0 1795 R
0 -63 V
1862 251 M
0 63 V
0 1795 R
0 -63 V
2208 251 M
0 63 V
0 1795 R
0 -63 V
2553 251 M
0 63 V
0 1795 R
0 -63 V
2899 251 M
0 63 V
0 1795 R
0 -63 V
3244 251 M
0 63 V
0 1795 R
0 -63 V
480 251 M
2937 0 V
0 1858 V
-2937 0 V
480 251 L
LT0
2328 1900 D
653 322 D
826 409 D
998 500 D
1171 595 D
826 407 D
1171 603 D
1517 809 D
1862 1030 D
998 496 D
1517 808 D
2035 1132 D
2553 1482 D
1171 589 D
1862 1025 D
2553 1474 D
3244 1962 D
2268 1900 M
180 0 V
-180 31 R
0 -62 V
180 62 R
0 -62 V
653 322 M
-31 0 R
62 0 V
-62 0 R
62 0 V
142 87 R
-31 0 R
62 0 V
-62 0 R
62 0 V
141 91 R
-31 0 R
62 0 V
-62 0 R
62 0 V
142 95 R
0 1 V
-31 -1 R
62 0 V
-62 1 R
62 0 V
826 407 M
-31 0 R
62 0 V
-62 0 R
62 0 V
314 196 R
0 1 V
-31 -1 R
62 0 V
-62 1 R
62 0 V
315 204 R
0 3 V
-31 -3 R
62 0 V
-62 3 R
62 0 V
314 217 R
0 5 V
-31 -5 R
62 0 V
-62 5 R
62 0 V
998 496 M
0 1 V
-31 -1 R
62 0 V
-62 1 R
62 0 V
488 309 R
0 3 V
-31 -3 R
62 0 V
-62 3 R
62 0 V
487 321 R
0 5 V
-31 -5 R
62 0 V
-62 5 R
62 0 V
487 343 R
0 9 V
-31 -9 R
62 0 V
-62 9 R
62 0 V
1171 589 M
0 1 V
-31 -1 R
62 0 V
-62 1 R
62 0 V
660 432 R
0 5 V
-31 -5 R
62 0 V
-62 5 R
62 0 V
660 442 R
0 10 V
-31 -10 R
62 0 V
-62 10 R
62 0 V
660 474 R
0 18 V
-31 -18 R
62 0 V
-62 18 R
62 0 V
stroke
grestore
end
showpage
}
\put(2208,1900){\makebox(0,0)[r]{$W(I,J,\lambda=0.1)/W(I,J,\lambda=0)$}}
\put(1700,1700){\makebox(0,0)[r]{$8^4 - \mbox{lattice}, \; \beta=6.0$}}
\put(1948,0){\makebox(0,0){$\mbox{Loop area} \; I \times J$}}
\put(3244,151){\makebox(0,0){16}}
\put(2899,151){\makebox(0,0){14}}
\put(2553,151){\makebox(0,0){12}}
\put(2208,151){\makebox(0,0){10}}
\put(1862,151){\makebox(0,0){8}}
\put(1517,151){\makebox(0,0){6}}
\put(1171,151){\makebox(0,0){4}}
\put(826,151){\makebox(0,0){2}}
\put(480,151){\makebox(0,0){0}}
\put(420,1940){\makebox(0,0)[r]{2}}
\put(420,1602){\makebox(0,0)[r]{1.8}}
\put(420,1264){\makebox(0,0)[r]{1.6}}
\put(420,927){\makebox(0,0)[r]{1.4}}
\put(420,589){\makebox(0,0)[r]{1.2}}
\put(420,251){\makebox(0,0)[r]{1}}
\end{picture}}
\caption{\label{lratios}Ratio of Wilson loops for chemical potentials
$\lambda=0.1$ and $\lambda=0.0$. The ratio increases stronger than
linearly with the size of the Wilson loops.}
\end{figure}

\begin{figure}
\centerline{\input{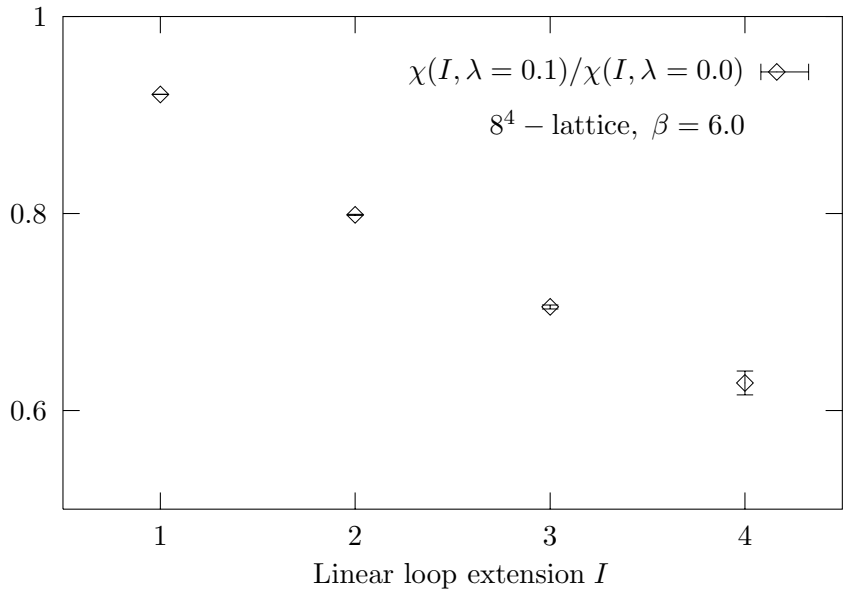}}
\caption{\label{cratios}Ratio of Creutz ratios for chemical potentials
$\lambda=0.1$ and $\lambda=0.0$ as a function of the linear extension
of the considered Wilson loops.}
\end{figure}

To investigate the influence of the current $J_{m,\mu}$ on physical
quantities we calculated Wilson loops and Creutz ratios on a
$8^4$-lattice at $\beta=6.0$ and compared the results for zero and
finite chemical potential $\lambda$. In fig.~\ref{lratios} the ratio
of Wilson loops obtained with $\lambda=0.1$ and $\lambda=0.0$
(standard Wilson action) is shown. It is clearly seen that a
suppression of magnetic currents manifests itself in a less disordered
lattice: For all sizes of Wilson loops the considered ratio is larger
than one and increases with increasing size of the loops. This means
that the suppression of magnetic currents (\ref{jnorm}) corresponding
to geometrical objects of the size of a unit cube especially
influences the behaviour of objects being large in terms of the
lattice spacing, i.e. large Wilson loops which determine the
non-perturbative properties of the theory. To get a more quantitative
insight we analyzed the behaviour of Creutz ratios $\chi$
\begin{equation}
\label{chir}
\chi(I,\lambda) = -\ln
\frac{W(I,I,\lambda)W(I-1,I-1,\lambda)}{W(I-1,I,\lambda)W(I,I-1,\lambda)}
\end{equation}
for the above discussed Wilson loops estimating the string
tension. Since the ratio of Wilson loops shown in fig.~\ref{lratios}
increases stronger than linearly, the ratio
$\chi(\lambda=0.1)/\chi(\lambda=0.0)$ decreases with the size of the
loops. This behaviour is shown in fig.~\ref{cratios}. The larger the
size of Wilson loops the stronger is the influence of the suppression
of magnetic currents on Creutz ratios. Taking the numerical results of
$\chi(4,\lambda)$ as an asymptotic value for the string tension
$\hat{\sigma}$ in lattice units we are able to estimate the decreasing
of the lattice spacing $a$
\begin{equation}
\label{lspacing}
a(\lambda=0.1) =
\sqrt{\frac{\hat{\sigma}(\lambda=0.1)}{\hat{\sigma}(\lambda=0.0)}} \;
a(\lambda=0.0) \;\; \approx \;\; 0.79 \; a(\lambda=0.0).
\end{equation}
We want to point out that for the chosen parameters $\beta=6.0$ and
$\lambda=0.1$ the average contribution of the extra term $\lambda \;
\sum_{x,\mu}  | J_{m,\mu} | $ to the total action (\ref{maction}) is
approximately $7 \%$. 
 
In addition to zero temperature investigations we also performed
numerical simulations at finite temperature to study the influence of
the magnetic current $J_{m,\mu}$ on the deconfinement phase
transition. The order parameter of the finite temperature phase
transition is the expectation value $\langle L \rangle$ of the
Polyakov loop
\begin{equation}
\label{polyakov}
L(\vec{x}) \; = \; \mbox{Tr} \prod_{t=1}^{N_\tau} U_4(\vec{x},t).
\end{equation}
In the thermodynamic limit $\langle L \rangle$ is expected to be zero
in the confined and non-zero in the deconfined phase. On a finite
lattice, however, $\langle L \rangle$ equals zero for all inverse
couplings $\beta$ because of the non-zero probability of tunneling
between different states related by the $Z(3)$-symmetry. We thus
measured the finite lattice ``order parameter'' $\langle |L| \rangle$
given by
\begin{equation}
\label{pollat}
\langle |L| \rangle \; = \; \langle | \frac{1}{N_{\sigma}^3}
\sum_{\vec{x}} L(\vec{x}) | \rangle,
\end{equation}
where the sum extends over the whole spatial lattice. We performed
runs on lattices with size $8^3 \times 4$ and $12^3 \times 4$ for
chemical potentials $\lambda=0.00$ (standard Wilson action) and
$\lambda = 0.05$. The results for $\langle |L| \rangle$ as a function
of the inverse coupling constant $\beta$ are shown in fig.~\ref{pol8}
and \ref{pol12}. It is clearly seen that a non-zero chemical potential
$\lambda$ shifts the phase transition towards a smaller critical value
of $\beta$. Hence, the addition of an extra term to the Wilson action
in (\ref{maction}) suppressing magnetic currents for $\lambda > 0$
corresponds to a system with a larger effective inverse coupling
constant. In other words, if configurations with large magnetic
currents $J_{m,\mu}$ are removed from the path integral, the phase
transition already occurs at smaller values of $\beta$, and it is more
favourable for the system to be in the $Z(3)$ broken than in the
$Z(3)$ symmetric phase. From fig.~\ref{pol8} and \ref{pol12} one can
estimate the shift of the phase transition. For $\lambda=0.05$ we find
that $\Delta\beta$ is approximately $-0.2$, whereas the current
contribution $\lambda \sum_{x,\mu} | J_{m,\mu} |$ to the total action
(\ref{maction}) is about $4 \%$ for $\beta \in [5.3,5.7]$.

\begin{figure}
\centerline{\input{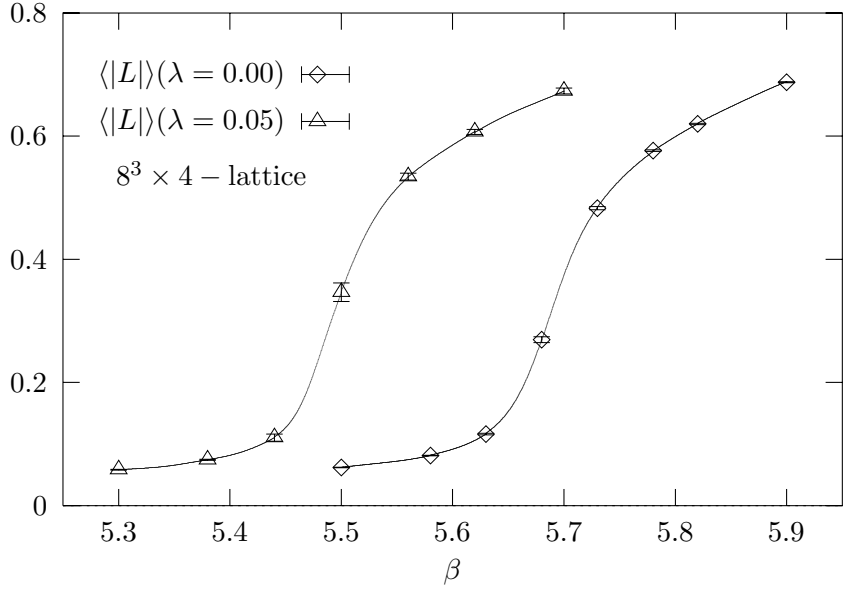}}
\caption{\label{pol8}The expectation value of the Polyakov loop as a
function of the inverse coupling $\beta$ for chemical potentials
$\lambda=0.00$ and $\lambda=0.05$ for a $8^3 \times 4$-lattice. In the
case of $\lambda \neq 0$ the critical $\beta$ is shifted towards a
smaller value. The symbols (diamonds for $\lambda=0.00$ and triangles
for $\lambda=0.05$) denote the numerical results of Monte Carlo runs
at different values of $\beta$ whereas the continuous curves were
computed with a multi-histogram analysis \cite{ferren,huang}.}
\end{figure}
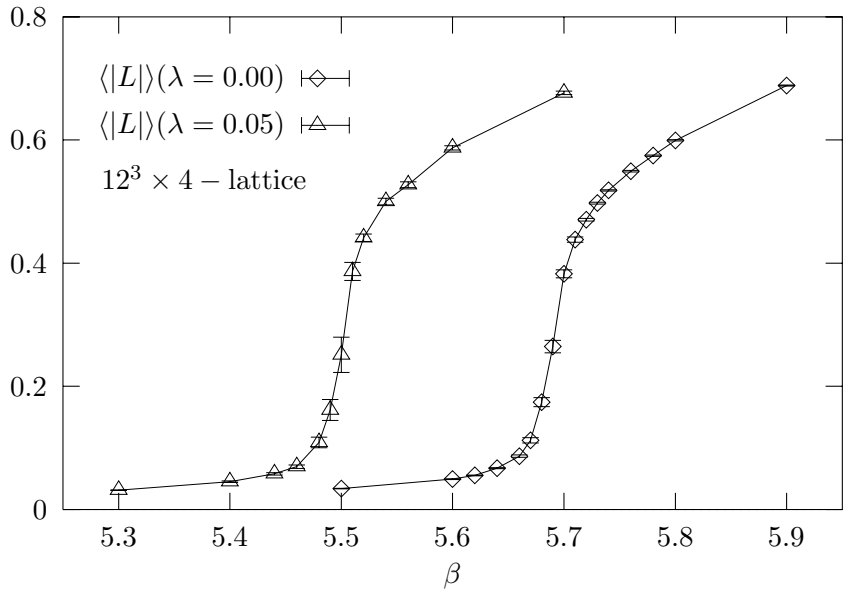
\begin{figure}
\centerline{
\setlength{\unitlength}{0.1bp}
\special{!
/gnudict 40 dict def
gnudict begin
/Color false def
/Solid false def
/gnulinewidth 2.000 def
/vshift -33 def
/dl {10 mul} def
/hpt 31.5 def
/vpt 31.5 def
/M {moveto} bind def
/L {lineto} bind def
/R {rmoveto} bind def
/V {rlineto} bind def
/vpt2 vpt 2 mul def
/hpt2 hpt 2 mul def
/Lshow { currentpoint stroke M
  0 vshift R show } def
/Rshow { currentpoint stroke M
  dup stringwidth pop neg vshift R show } def
/Cshow { currentpoint stroke M
  dup stringwidth pop -2 div vshift R show } def
/DL { Color {setrgbcolor Solid {pop []} if 0 setdash }
 {pop pop pop Solid {pop []} if 0 setdash} ifelse } def
/BL { stroke gnulinewidth 2 mul setlinewidth } def
/AL { stroke gnulinewidth 2 div setlinewidth } def
/PL { stroke gnulinewidth setlinewidth } def
/LTb { BL [] 0 0 0 DL } def
/LTa { AL [1 dl 2 dl] 0 setdash 0 0 0 setrgbcolor } def
/LT0 { PL [] 0 1 0 DL } def
/LT1 { PL [4 dl 2 dl] 0 0 1 DL } def
/LT2 { PL [2 dl 3 dl] 1 0 0 DL } def
/LT3 { PL [1 dl 1.5 dl] 1 0 1 DL } def
/LT4 { PL [5 dl 2 dl 1 dl 2 dl] 0 1 1 DL } def
/LT5 { PL [4 dl 3 dl 1 dl 3 dl] 1 1 0 DL } def
/LT6 { PL [2 dl 2 dl 2 dl 4 dl] 0 0 0 DL } def
/LT7 { PL [2 dl 2 dl 2 dl 2 dl 2 dl 4 dl] 1 0.3 0 DL } def
/LT8 { PL [2 dl 2 dl 2 dl 2 dl 2 dl 2 dl 2 dl 4 dl] 0.5 0.5 0.5 DL } def
/P { stroke [] 0 setdash
  currentlinewidth 2 div sub M
  0 currentlinewidth V stroke } def
/D { stroke [] 0 setdash 2 copy vpt add M
  hpt neg vpt neg V hpt vpt neg V
  hpt vpt V hpt neg vpt V closepath stroke
  P } def
/A { stroke [] 0 setdash vpt sub M 0 vpt2 V
  currentpoint stroke M
  hpt neg vpt neg R hpt2 0 V stroke
  } def
/B { stroke [] 0 setdash 2 copy exch hpt sub exch vpt add M
  0 vpt2 neg V hpt2 0 V 0 vpt2 V
  hpt2 neg 0 V closepath stroke
  P } def
/C { stroke [] 0 setdash exch hpt sub exch vpt add M
  hpt2 vpt2 neg V currentpoint stroke M
  hpt2 neg 0 R hpt2 vpt2 V stroke } def
/T { stroke [] 0 setdash 2 copy vpt 1.12 mul add M
  hpt neg vpt -1.62 mul V
  hpt 2 mul 0 V
  hpt neg vpt 1.62 mul V closepath stroke
  P  } def
/S { 2 copy A C} def
end
}
\begin{picture}(3600,2160)(0,0)
\special{"
gnudict begin
gsave
50 50 translate
0.100 0.100 scale
0 setgray
/Helvetica findfont 100 scalefont setfont
newpath
-500.000000 -500.000000 translate
LTa
480 251 M
2937 0 V
LTb
480 251 M
63 0 V
2874 0 R
-63 0 V
480 716 M
63 0 V
2874 0 R
-63 0 V
480 1180 M
63 0 V
2874 0 R
-63 0 V
480 1645 M
63 0 V
2874 0 R
-63 0 V
480 2109 M
63 0 V
2874 0 R
-63 0 V
690 251 M
0 63 V
0 1795 R
0 -63 V
1109 251 M
0 63 V
0 1795 R
0 -63 V
1529 251 M
0 63 V
0 1795 R
0 -63 V
1948 251 M
0 63 V
0 1795 R
0 -63 V
2368 251 M
0 63 V
0 1795 R
0 -63 V
2788 251 M
0 63 V
0 1795 R
0 -63 V
3207 251 M
0 63 V
0 1795 R
0 -63 V
480 251 M
2937 0 V
0 1858 V
-2937 0 V
480 251 L
LT0
1439 1877 D
1529 330 D
1948 366 D
2032 380 D
2116 407 D
2200 452 D
2242 512 D
2284 656 D
2326 866 D
2368 1140 D
2410 1269 D
2452 1344 D
2494 1407 D
2536 1455 D
2620 1527 D
2704 1586 D
2788 1643 D
3207 1850 D
1379 1877 M
180 0 V
-180 31 R
0 -62 V
180 62 R
0 -62 V
1529 329 M
0 2 V
-31 -2 R
62 0 V
-62 2 R
62 0 V
388 33 R
0 3 V
-31 -3 R
62 0 V
-62 3 R
62 0 V
53 11 R
0 3 V
-31 -3 R
62 0 V
-62 3 R
62 0 V
53 24 R
0 4 V
-31 -4 R
62 0 V
-62 4 R
62 0 V
53 39 R
0 9 V
-31 -9 R
62 0 V
-62 9 R
62 0 V
11 45 R
0 20 V
-31 -20 R
62 0 V
-62 20 R
62 0 V
11 117 R
0 34 V
-31 -34 R
62 0 V
-62 34 R
62 0 V
11 169 R
0 47 V
-31 -47 R
62 0 V
-62 47 R
62 0 V
11 236 R
0 30 V
-31 -30 R
62 0 V
-62 30 R
62 0 V
11 104 R
0 20 V
-31 -20 R
62 0 V
-62 20 R
62 0 V
11 60 R
0 10 V
-31 -10 R
62 0 V
-62 10 R
62 0 V
11 54 R
0 7 V
-31 -7 R
62 0 V
-62 7 R
62 0 V
11 42 R
0 5 V
-31 -5 R
62 0 V
-62 5 R
62 0 V
53 67 R
0 6 V
-31 -6 R
62 0 V
-62 6 R
62 0 V
53 53 R
0 5 V
-31 -5 R
62 0 V
-62 5 R
62 0 V
53 53 R
0 5 V
-31 -5 R
62 0 V
-62 5 R
62 0 V
388 202 R
0 4 V
-31 -4 R
62 0 V
-62 4 R
62 0 V
LT0
1529 330 M
419 36 V
84 14 V
84 27 V
84 45 V
42 60 V
42 144 V
42 210 V
42 274 V
42 129 V
42 75 V
42 63 V
42 48 V
84 72 V
84 59 V
84 57 V
419 207 V
1439 1700 T
690 324 T
1109 356 T
1277 386 T
1361 413 T
1445 504 T
1487 626 T
1529 834 T
1571 1149 T
1613 1276 T
1697 1413 T
1781 1476 T
1948 1615 T
2368 1822 T
1379 1700 M
180 0 V
-180 31 R
0 -62 V
180 62 R
0 -62 V
690 323 M
0 2 V
-31 -2 R
62 0 V
-62 2 R
62 0 V
388 28 R
0 6 V
-31 -6 R
62 0 V
-62 6 R
62 0 V
137 22 R
0 9 V
-31 -9 R
62 0 V
-62 9 R
62 0 V
53 16 R
0 13 V
-31 -13 R
62 0 V
-62 13 R
62 0 V
53 65 R
0 40 V
-31 -40 R
62 0 V
-62 40 R
62 0 V
11 63 R
0 79 V
-31 -79 R
62 0 V
-62 79 R
62 0 V
11 102 R
0 133 V
1498 768 M
62 0 V
-62 133 R
62 0 V
11 214 R
0 68 V
-31 -68 R
62 0 V
-62 68 R
62 0 V
11 79 R
0 28 V
-31 -28 R
62 0 V
-62 28 R
62 0 V
53 110 R
0 25 V
-31 -25 R
62 0 V
-62 25 R
62 0 V
53 40 R
0 22 V
-31 -22 R
62 0 V
-62 22 R
62 0 V
136 120 R
0 16 V
-31 -16 R
62 0 V
-62 16 R
62 0 V
389 193 R
0 13 V
-31 -13 R
62 0 V
-62 13 R
62 0 V
LT0
690 324 M
419 32 V
168 30 V
84 27 V
84 91 V
42 122 V
42 208 V
42 315 V
42 127 V
84 137 V
84 63 V
167 139 V
420 207 V
stroke
grestore
end
showpage
}
\put(1319,1700){\makebox(0,0)[r]{$\langle |L| \rangle(\lambda=0.05)$}}
\put(1319,1877){\makebox(0,0)[r]{$\langle |L| \rangle(\lambda=0.00)$}}
\put(1400,1500){\makebox(0,0)[r]{$12^3 \times 4 - \mbox{lattice}$}}
\put(1948,0){\makebox(0,0){$\beta$}}
\put(3207,151){\makebox(0,0){5.9}}
\put(2788,151){\makebox(0,0){5.8}}
\put(2368,151){\makebox(0,0){5.7}}
\put(1948,151){\makebox(0,0){5.6}}
\put(1529,151){\makebox(0,0){5.5}}
\put(1109,151){\makebox(0,0){5.4}}
\put(690,151){\makebox(0,0){5.3}}
\put(420,2109){\makebox(0,0)[r]{0.8}}
\put(420,1645){\makebox(0,0)[r]{0.6}}
\put(420,1180){\makebox(0,0)[r]{0.4}}
\put(420,716){\makebox(0,0)[r]{0.2}}
\put(420,251){\makebox(0,0)[r]{0}}
\end{picture}}
\caption{\label{pol12}The same as in fig.~\ref{pol8} for a $12^3
\times 4$-lattice except that the curves are linear interpolations
between the results of Monte Carlo runs drawn to guide the eye.}
\end{figure}
Besides the Polyakov loop we also measured the distribution of the
real part of the plaquette trace. In fig.~\ref{plaq} we show a
comparison of plaquette distributions on the $12^3 \times 4$-lattice
at $\beta=5.6$ for $\lambda=0.00$ and $\lambda=0.05$. The
corresponding numerical values of $\langle |L| \rangle$ are displayed
in fig.~\ref{pol12}. As expected the suppression of magnetic currents
causes the plaquettes to be closer to the unit matrix. Plaquette
distributions corresponding to the same value of $\langle |L| \rangle$
but different values of $\beta$ for $\lambda=0.00$ and $\lambda=0.05$
cannot be distinguished.

By choosing the chemical potential $\lambda$ large enough the
confinement phase can be completely removed for positive values of
$\beta$. Starting from an ordered lattice and choosing $\lambda=10.0$
the system is not able to evolve from the deconfined into the
disordered confined phase, also not for $\beta=0.0$, see
fig.~\ref{pol_l=10}. This shows that for the considered lattice sizes
$8^3 \times 4$ and $12^3 \times 4$ there is no confined phase without
presence of magnetic currents $J_{m,\mu}$ for positive values of
$\beta$.
\begin{figure}
\centerline{
\setlength{\unitlength}{0.1bp}
\special{!
/gnudict 40 dict def
gnudict begin
/Color false def
/Solid false def
/gnulinewidth 2.000 def
/vshift -33 def
/dl {10 mul} def
/hpt 31.5 def
/vpt 31.5 def
/M {moveto} bind def
/L {lineto} bind def
/R {rmoveto} bind def
/V {rlineto} bind def
/vpt2 vpt 2 mul def
/hpt2 hpt 2 mul def
/Lshow { currentpoint stroke M
  0 vshift R show } def
/Rshow { currentpoint stroke M
  dup stringwidth pop neg vshift R show } def
/Cshow { currentpoint stroke M
  dup stringwidth pop -2 div vshift R show } def
/DL { Color {setrgbcolor Solid {pop []} if 0 setdash }
 {pop pop pop Solid {pop []} if 0 setdash} ifelse } def
/BL { stroke gnulinewidth 2 mul setlinewidth } def
/AL { stroke gnulinewidth 2 div setlinewidth } def
/PL { stroke gnulinewidth setlinewidth } def
/LTb { BL [] 0 0 0 DL } def
/LTa { AL [1 dl 2 dl] 0 setdash 0 0 0 setrgbcolor } def
/LT0 { PL [] 0 1 0 DL } def
/LT1 { PL [4 dl 2 dl] 0 0 1 DL } def
/LT2 { PL [2 dl 3 dl] 1 0 0 DL } def
/LT3 { PL [1 dl 1.5 dl] 1 0 1 DL } def
/LT4 { PL [5 dl 2 dl 1 dl 2 dl] 0 1 1 DL } def
/LT5 { PL [4 dl 3 dl 1 dl 3 dl] 1 1 0 DL } def
/LT6 { PL [2 dl 2 dl 2 dl 4 dl] 0 0 0 DL } def
/LT7 { PL [2 dl 2 dl 2 dl 2 dl 2 dl 4 dl] 1 0.3 0 DL } def
/LT8 { PL [2 dl 2 dl 2 dl 2 dl 2 dl 2 dl 2 dl 4 dl] 0.5 0.5 0.5 DL } def
/P { stroke [] 0 setdash
  currentlinewidth 2 div sub M
  0 currentlinewidth V stroke } def
/D { stroke [] 0 setdash 2 copy vpt add M
  hpt neg vpt neg V hpt vpt neg V
  hpt vpt V hpt neg vpt V closepath stroke
  P } def
/A { stroke [] 0 setdash vpt sub M 0 vpt2 V
  currentpoint stroke M
  hpt neg vpt neg R hpt2 0 V stroke
  } def
/B { stroke [] 0 setdash 2 copy exch hpt sub exch vpt add M
  0 vpt2 neg V hpt2 0 V 0 vpt2 V
  hpt2 neg 0 V closepath stroke
  P } def
/C { stroke [] 0 setdash exch hpt sub exch vpt add M
  hpt2 vpt2 neg V currentpoint stroke M
  hpt2 neg 0 R hpt2 vpt2 V stroke } def
/T { stroke [] 0 setdash 2 copy vpt 1.12 mul add M
  hpt neg vpt -1.62 mul V
  hpt 2 mul 0 V
  hpt neg vpt 1.62 mul V closepath stroke
  P  } def
/S { 2 copy A C} def
end
}
\begin{picture}(2339,1511)(0,0)
\special{"
gnudict begin
gsave
50 50 translate
0.100 0.100 scale
0 setgray
/Helvetica findfont 100 scalefont setfont
newpath
-500.000000 -500.000000 translate
LTa
LTb
480 151 M
63 0 V
1613 0 R
-63 0 V
480 478 M
63 0 V
1613 0 R
-63 0 V
480 806 M
63 0 V
1613 0 R
-63 0 V
480 1133 M
63 0 V
1613 0 R
-63 0 V
480 1460 M
63 0 V
1613 0 R
-63 0 V
666 151 M
0 63 V
0 1246 R
0 -63 V
1039 151 M
0 63 V
0 1246 R
0 -63 V
1411 151 M
0 63 V
0 1246 R
0 -63 V
1784 151 M
0 63 V
0 1246 R
0 -63 V
2156 151 M
0 63 V
0 1246 R
0 -63 V
480 151 M
1676 0 V
0 1309 V
-1676 0 V
480 151 L
LT0
1140 1150 M
180 0 V
489 151 M
19 0 V
19 0 V
18 0 V
19 0 V
18 0 V
19 0 V
19 0 V
18 0 V
19 1 V
19 0 V
18 1 V
19 0 V
18 1 V
19 1 V
19 1 V
18 1 V
19 1 V
19 2 V
18 2 V
19 2 V
18 2 V
19 3 V
19 3 V
18 3 V
19 4 V
18 5 V
19 5 V
19 6 V
18 6 V
19 7 V
19 8 V
18 9 V
19 10 V
18 11 V
19 12 V
19 13 V
18 15 V
19 15 V
19 18 V
18 19 V
19 20 V
18 22 V
19 24 V
19 26 V
18 27 V
19 29 V
19 31 V
18 33 V
19 35 V
18 37 V
19 38 V
19 40 V
18 41 V
19 43 V
19 43 V
18 45 V
19 44 V
18 45 V
19 44 V
19 43 V
18 41 V
19 40 V
19 36 V
18 33 V
19 29 V
18 22 V
19 18 V
19 9 V
18 2 V
19 -5 V
18 -15 V
19 -26 V
19 -35 V
18 -45 V
19 -55 V
19 -66 V
18 -75 V
19 -83 V
18 -88 V
19 -94 V
19 -94 V
18 -93 V
19 -88 V
19 -80 V
18 -68 V
19 -53 V
18 -37 V
19 -20 V
19 -6 V
LT2
1140 1300 M
180 0 V
489 151 M
19 0 V
19 0 V
18 0 V
19 0 V
18 0 V
19 0 V
19 1 V
18 0 V
19 1 V
19 2 V
18 2 V
19 2 V
18 1 V
19 2 V
19 3 V
18 2 V
19 3 V
19 3 V
18 4 V
19 4 V
18 4 V
19 5 V
19 5 V
18 6 V
19 6 V
18 8 V
19 7 V
19 9 V
18 9 V
19 10 V
19 11 V
18 13 V
19 13 V
18 14 V
19 15 V
19 17 V
18 17 V
19 19 V
19 20 V
18 22 V
19 23 V
18 25 V
19 25 V
19 28 V
18 28 V
19 29 V
19 33 V
18 33 V
19 33 V
18 35 V
19 36 V
19 35 V
18 40 V
19 36 V
19 37 V
18 39 V
19 35 V
18 35 V
19 35 V
19 30 V
18 30 V
19 26 V
19 22 V
18 21 V
19 13 V
18 9 V
19 1 V
19 -3 V
18 -10 V
19 -17 V
18 -26 V
19 -35 V
19 -41 V
18 -50 V
19 -58 V
19 -63 V
18 -68 V
19 -78 V
18 -77 V
19 -79 V
19 -79 V
18 -76 V
19 -70 V
19 -63 V
18 -52 V
19 -40 V
18 -27 V
19 -15 V
19 -5 V
stroke
grestore
end
showpage
}
\put(1100,1300){\makebox(0,0)[r]{$\lambda=0.00$}}
\put(1100,1150){\makebox(0,0)[r]{$\lambda=0.05$}}
\put(2156,51){\makebox(0,0){3}}
\put(1784,51){\makebox(0,0){2}}
\put(1411,51){\makebox(0,0){1}}
\put(1600,-100){\makebox(0,0)[r]{$\mbox{Re Tr} U_{\Box}$}}
\put(1039,51){\makebox(0,0){0}}
\put(666,51){\makebox(0,0){-1}}
\put(420,1460){\makebox(0,0)[r]{0.04}}
\put(420,1133){\makebox(0,0)[r]{0.03}}
\put(420,806){\makebox(0,0)[r]{0.02}}
\put(420,478){\makebox(0,0)[r]{0.01}}
\put(420,151){\makebox(0,0)[r]{0}}
\end{picture}}
\vspace{0.5cm}
\caption{\label{plaq}Distributions of the real part of the plaquette
trace for $\beta=5.6$ and the $\lambda$- values $0.00$ (dashed line)
and $0.05$ (solid line). According to fig.~\ref{pol12} the lattice is
in the confined phase for $\lambda = 0.00$ whereas it is in the
deconfined phase for $\lambda = 0.05$.}
\end{figure}
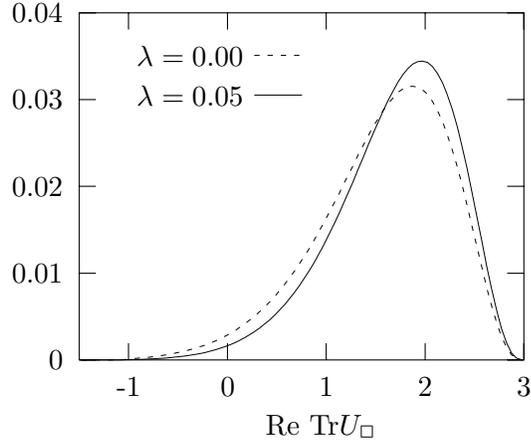
\begin{figure}
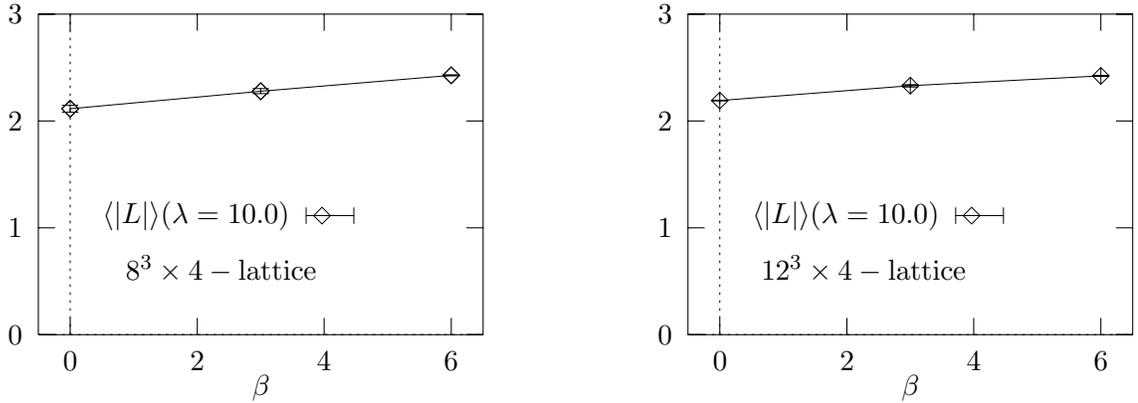

\centerline{\input{pol_l=10_8.tex}\input{pol_l=10_12.tex}}
\caption{\label{pol_l=10}The expectation value of the Polyakov loop as
a function of the inverse coupling $\beta$ for chemical potential
$\lambda=10.0$ for a $8^3 \times 4$-lattice (left plot) and a $12^3
\times 4$-lattice (right plot). At this value of $\lambda$ there is no
confined phase left for positive values of the inverse coupling
$\beta$.}
\end{figure}

\section{Conclusion and discussion}

In this letter we studied the influence of colour magnetic currents
(\ref{mcur}) originally defined in ref.~\cite{skala} on the confining
properties of $SU(3)$ lattice gauge theory. We introduced a chemical
potential $\lambda$ and added an extra term to the standard Wilson
action suppressing large magnetic currents. With this modified action
we performed numerical simulations at zero and finite temperature
lattices. We found that a partial absence of magnetic currents leads
to a drastic decrease in the confining string tension. At finite
temperature the suppression of magnetic currents shifts the phase
transition towards smaller values of the inverse coupling constant
$\beta$. Additionally, for the considered lattice sizes it was shown
that there exists a finite chemical potential $\lambda$ for which the
confined phase is completely removed for positive values of
$\beta$. We conclude that the magnetic currents strongly influence the
confining properties of our considered $SU(3)$ lattice gauge
theory. However, an interpretation of these currents in terms of
physical objects is not directly possible since we do not observe
scaling behaviour of the current density (\ref{jma}). As already
mentioned above, the reason for this might be that there are only
positive contributions to the expectation value $\langle | J_m |
\rangle$ and that fluctuations which are not of topological origin do
not cancel. According to our opinion, on a lattice with finite spacing
there are in general two different types of contributions to the magnetic
current $J_{m,\mu}$. One type of contributions is due to the finite
lattice spacing. They can be regarded as lattice artifacts and vanish
in the continuum. Hence, they will not contribute to the continuum
limit of the expectation value $\langle | J_m | \rangle$. The
other type of contributions is of topological origin. They belong to
gauge field configurations which become singular in particular gauges
in the continuum and thus could guarantee that the operator
$J_{m,\mu}$ does not become trivial in the continuum limit. One
candidate for such configurations is for instance the Euclidean pure
gauge theory analogue of the 't~Hooft-Polyakov-like monopole in the
BPS limit. That this monopole configuration could play an important
role in the dual superconductor picture of confinement was discussed in
ref.~\cite{vdsijs1}. Moreover, in ref.~\cite{vdsijs2} it was argued
that in an appropriate gauge such a monopole configuration becomes singular.
Analogous arguments should be valid for another type of gauge field
configurations, magnetic vortices. These were suggested to play a
crucial role in the confinement mechanism \cite{vortex}, an idea which gained
support from recent lattice calculations \cite{green}.

To summarize, we believe that the topological contributions to the
magnetic currents strongly influence the confining properties of
QCD. Unfortunately, the magnetic current operator is not able to
identify its topological sources which determine the behaviour of its
expectation value in the continuum. Thus, we can only speculate about
this question. But for cubes of finite size, we know that the
topological objects lead to magnetic currents behaving as predicted by
the dual superconductor picture.
  
\section*{Acknowledgments}
We thank M.~Ilgenfritz for providing us with a pseudo heatbath algorithm.

\end{document}